\documentclass{caosp309}

\usepackage{graphicx}
\usepackage{natbib}
\usepackage{upgreek}
\usepackage{amssymb}
\articleNo{301}
\pubyear{2020}
\volume{50}
\volnumber{3}
\firstpage{1}
\received{May 1, 2020}
\accepted{July 28, 2020}

\begin{document}

%
\hauthor{Simon J. Murphy}

\title{Pulsations in binary stars -- a short review}
\author{
        Simon J. Murphy\inst{1}\orcid{0000-0002-5648-3107}
       }

%
\institute{
Centre for Astrophysics, University of Southern Queensland, Toowoomba, QLD 4350, Australia, \email{simon.murphy@unisq.edu.au}\\
          }

\date{March 8, 2003}

\maketitle

\begin{abstract}
Asteroseismology has become an indispensable method for measuring stellar ages and radii, while binary systems remain the most prevalent tool for determining stellar masses. The synergy of the two, namely pulsating stars in binary systems, offer even more than the sum of their parts. The sometimes-overwhelming number of pulsation models to be examined for asteroseismic modelling can be tightly constrained when dynamical masses for both components are available. Binaries offer twice the opportunity to measure an asteroseismic age, which is then applicable to both components of a system, often providing ages for stars that are otherwise very difficult to determine. Some stellar physics, such as the strength of internal mixing or of core overshoot, is so difficult to infer that only in binary systems do sufficient constraints exist to advance our asteroseismic models. The pulsations themselves can also be used to infer dynamical masses through pulsation timing, while eclipsing binaries provide opportunities to test asteroseismic scaling relations. In these proceedings of an invited talk I selectively review the synergies afforded by pulsations in binaries, consisting of a variety of main-sequence primary spectral types, to inferences made from the oscillations of red giants, even when their companions are undetectable. I also describe how the field is changing as we become more data rich through each generation of large photometric survey.
\end{abstract}

%
\section{Introduction}
\label{sec:intro}
Most stellar properties exhibit great variation across spectral types. Temperature and mass are obvious examples, but rotation rates and binary fraction are important yet often under-appreciated additions. Along the main sequence alone, rotation periods range from under 1\,d for some B and A stars, to over 30\,d for late-type dwarfs \citep{reinhold&hekker2020}. Meanwhile, the binary fraction drops by a factor of 10--15 from B primaries to M primaries \citep{moe&distefano2017,offneretal2023}\footnote{Strongly dependent on orbital period.}. The implications of these two effects are substantial when one considers that the majority of new binary stars were being found spectroscopically until recently \citep{halbwachsetal2023,holletal2023}.

The rotational broadening of spectral lines makes identification of binaries much harder for many reasons. When lines are very broad, it can become challenging to distinguish all but the strongest lines from the continuum. It is therefore almost impossible to tell when you are dealing with lines from two stars (a double-lined spectroscopic binary, or `SB2'), rather than from one. Further, if those spectral lines can be detected for even one of the stars, measuring their centres is also challenging (cf. \citealt{rucinski1999}), so it is not trivial to detect any variations in radial velocity (as a single-lined spectroscopic binary, or `SB1'). To further complicate matters, stars of B and A spectral types often pulsate \citep{kurtz2022,hey&aerts2024,murphyetal2024}, with typical pulsation velocities exceeding 1\,km\,s$^{-1}$ \citep{kjeldsen&bedding1995}. Those pulsations can induce radial velocity shifts and also spectral line profile variations (e.g. \citealt{rieutordetal2023,pollardetal2024}), further obfuscating any spectroscopic signature of a binary companion. Clearly, a complete survey of binarity amongst these stars will require non-spectroscopic techniques.

The present era has seen a photometric revolution with the launch of space-based missions such as CoRoT, {\it Kepler} and TESS, as well as ground-based surveys such as ASAS and KELT. Although motivated by the search for transiting planets, these missions have detected eclipsing binaries (EBs) with ease and in large numbers. The observing strategy is also perfect for asteroseismology -- the study of stellar pulsations (see \citealt{bowman&bugnet2024} for an introductory review). It is no surprise, then, that the all-sky observing strategy of TESS has picked up countless pulsating stars in EBs \citep{southworth&bowman2022,kahramanalicavusetal2022,kahramanalicavusetal2023,eze&handler2024}. These will be discussed later in this review. The long and almost uninterrupted time-series from {\it Kepler} and to some extent from TESS have also facilitated the detection of many pulsation timing binaries \citep{murphyetal2018,dholakiaetal2024}, which will also be covered in this review. The prospects for discovering more pulsators in binaries look good for both methods, with the continued operation of TESS and with the upcoming Roman Space Telescope \citep{huberetal2023}, PLATO mission \citep{aerts&tkachenko2024}, and Earth 2.0 (ET) telescope \citep{geetal2022}.

A similar short review on pulsating stars in binary systems was written as a conference proceedings in 2018 \citep{murphy2018}. The intention is for the present review to be synergistic with the earlier one and to provide updates in the field since then. For instance, the 2018 review covered compact pulsators in detail, and while TESS observations of compact pulsators have been published (e.g. \citealt{bognaretal2020,baranetal2023}), rather few reports have appeared of binaries amongst them (cf. \citealt{leeetal2022}). Compact objects therefore do not feature in this review (they were not part of the talk upon which this review is based, either). Early-type pulsators in EBs, on the other hand, have had a TESS renaissance, and so receive substantial coverage here. A particularly significant development since 2018 has been the discovery of tidally trapped pulsations \citep{handleretal2020,fulleretal2020}, which are also extensively reviewed. Nonetheless, there is naturally some overlap with the earlier review, particularly pertaining to pulsation timing but also to solar-like oscillators in binaries, which were cornerstones of the review talks then and now.

This review is laid out as follows. A few exemplary asteroseismic analyses of main-sequence solar-like oscillators in binaries are presented in Sec.\,\ref{sec:slo}, and their post-main-sequence counterparts are reviewed in Sec.\,\ref{sec:rgs}. A comprehensive update on pulsation timing over the past six years is provided in Sec.\,\ref{sec:TDs} along with a look at future prospects of this technique. Sec.\,\ref{sec:tidal} explores the newly described class of tidally trapped pulsations, where the effect of tides on the pulsations can be so great that the tidal axis becomes the pulsation axis. Finally, Sec.\,\ref{sec:EBs} describes the surge of discoveries of early-type pulsators in eclipsing binaries, and lessons that can be learned from such studies.

\section{Main-sequence solar-like oscillators in binaries}
\label{sec:slo}

\subsection{Photometry}

The \textit{Kepler} LEGACY sample contains 66 well-characterised main-sequence solar-like oscillators \citep{lundetal2017}, and many more such oscillators with more complicated power spectra were also observed. The detection of stochastic oscillations in main-sequence stars requires high cadence observations and long time-series. With \textit{Kepler}, this means short-cadence mode (60-s) observations, and the LEGACY sample only included stars with at least 1\,yr of data. TESS, by comparison, had a less-short cadence of 120\,s\footnote{TESS now has many different cadences available, from the 20-s `ultra-short' cadence to the original 1800-s full frame images (FFIs). The FFI cadence has now dropped to 200\,s.}, and an overall redder bandpass, in which the oscillation amplitudes appear reduced. The all-sky sampling strategy also disfavours the collation of longer time-series except at the ecliptic poles. Nonetheless, a few dozen main-sequence solar-like oscillators were also found with TESS \citep{hattetal2023}, along with many more subgiants and giants. Naturally, some of these were known spectroscopic binaries, although none were reported to have two sets of oscillations in the same light curve (i.e. none were double-pulsator binaries, or PB2s\footnote{named in analogy to the SB2s of spectroscopy}).

PB2s are promising from an asteroseismic perspective. The 2018 review provided multiple examples of solar-like oscillators in PB2s \citep{appourchauxetal2015,whiteetal2017a,ylietal2018} and explained their significance. In short, such systems should have the same age and metallicity, meaning that the frequency content of their combined power spectrum is governed largely by their difference in mass. This makes them valuable tests of stellar evolution and pulsation models. However, stars with mass ratios close to unity, such that the luminosity of one doesn't pale in comparison to the other, are predicted to be rare \citep{moe&distefano2017}, and it appears that we also detect far fewer than are predicted to exist (\citealt{miglioetal2014}, see discussion in \citealt{murphy2018}). New simulations that include {\it Gaia} data as priors are in preparation \citep{mazzietal2024}, and there is hope that the faster cadence (25\,s) of the PLATO mission will yield many more PB2s.

\subsection{Radial velocities}
\begin{figure}
\begin{center}
\includegraphics[width=1.0\linewidth,angle=0]{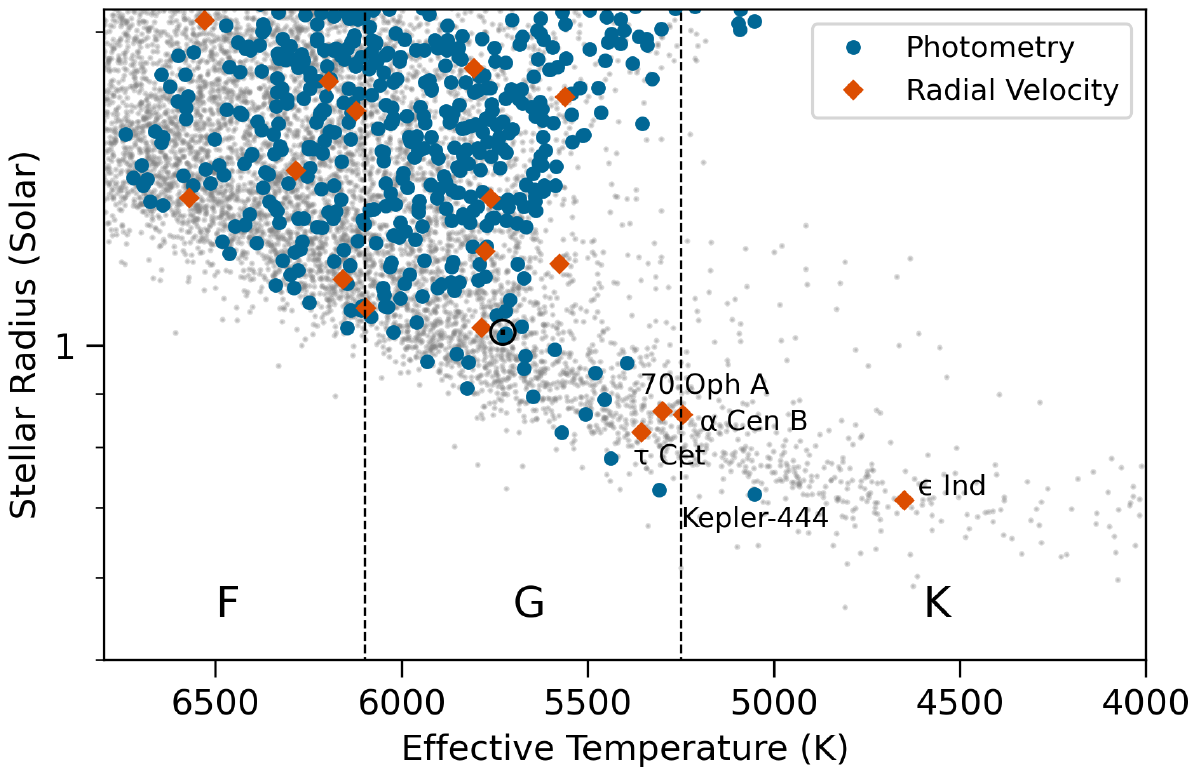}\\
\caption{Stellar radius--effective temperature diagram highlighting seismic detections from {\it Kepler} and TESS photometry (blue circles; \citealt{mathuretal2017,hattetal2023}), and radial-velocity campaigns (red diamonds; see e.g., \citealt{arentoftetal2008,kjeldsenetal2008a}, and references therein). The stellar background sample (grey dots) is taken from the TESS Input Catalog (TIC; \citealt{stassunetal2019}). The Sun is represented by its usual symbol. Approximate spectral type ranges (F, G, and K) are delimited by the vertical dashed lines. $\epsilon$\,Indi (K5\,V) is the coolest seismic dwarf observed to date (its interferometric radius and effective temperature were used to place it in the diagram; \citealt{rainsetal2020}). Figure and caption reproduced from \citet{campanteetal2024}.}
\label{fig:slo_detections} 
\end{center}
\end{figure}

Asteroseismology of solar-like oscillators in velocity space (as opposed to photometry) was not covered in the 2018 review. There are pros and cons to using radial velocities \citep{bedding&kjeldsen2007}. The signal to noise ratio in velocity observations is usually better, because the granulation background is much smaller. In principle, this means that for cooler stars, where convection is more vigorous, velocities are more likely to yield a detection. The drawback is that single-target, ground-based velocity campaigns are more challenging to organise and have stringent technical requirements: it takes a short exposure to sample the oscillations, and so the spectrograph must be capable of a fast readout. Meanwhile, it takes a large aperture to gather enough photons in a short exposure. Hence, while rather few instruments meet the technical requirements, the capabilities can even exceed space-based photometers. Fig.\,\ref{fig:slo_detections} shows the known solar-like oscillators from both photometry and radial velocities.

\begin{figure}
\begin{center}
\includegraphics[width=0.8\linewidth,angle=0]{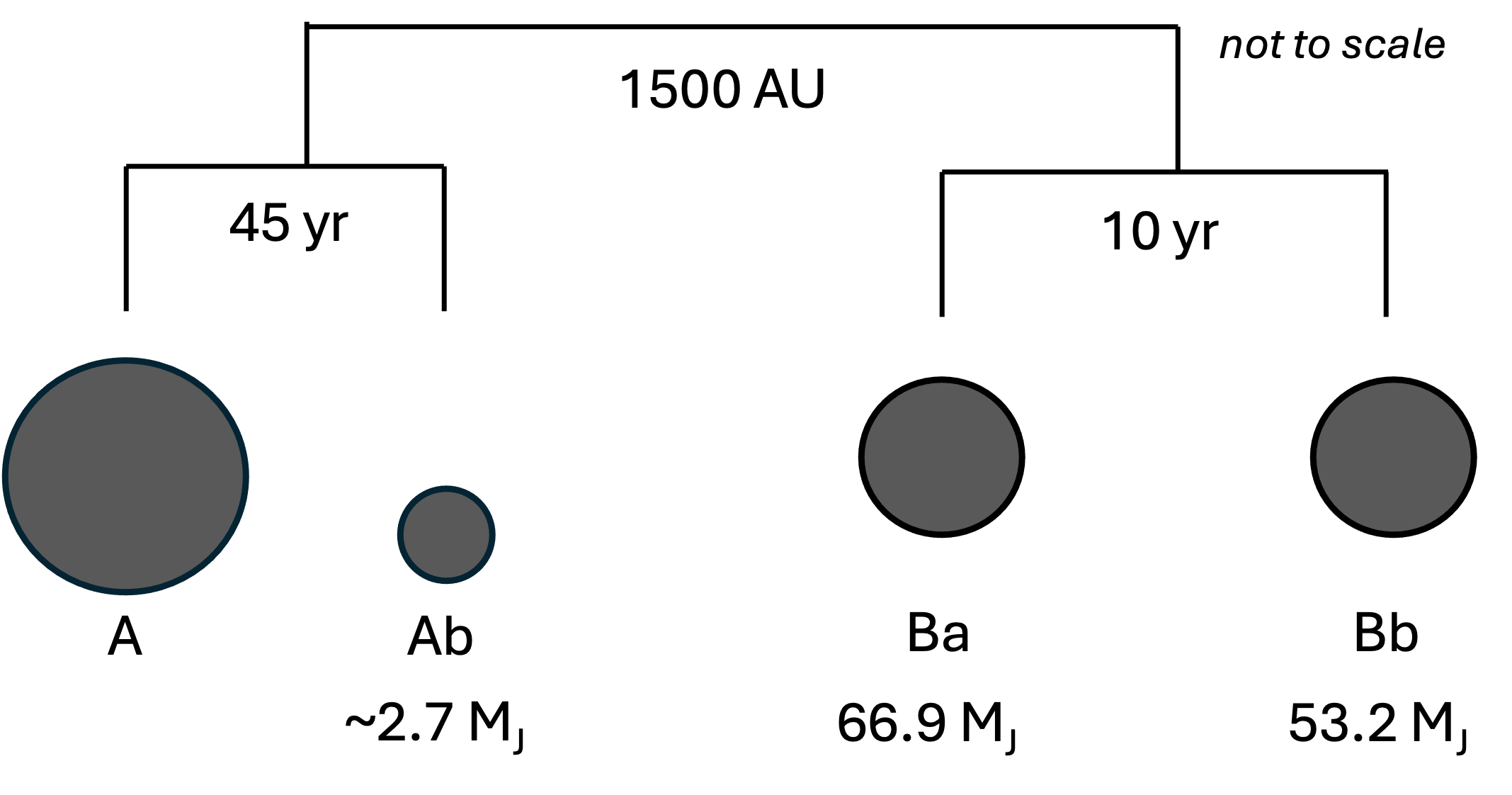}\\
\caption{Mobile diagram for the $\epsilon$\,Ind system. System properties are from \citet{lundkvistetal2024}, \citet{campanteetal2024}, and references therein.}
\label{fig:eps_ind} 
\end{center}
\end{figure}

The coolest main-sequence star in which oscillations have been detected is $\epsilon$\,Indi\,A, which is an interesting case in the context of pulsations in binaries because the system is multiple (Fig.\,\ref{fig:eps_ind}). The companions $\epsilon$\,Ind\,Ba and Bb are among the best studied brown dwarfs, while $\epsilon$\,Ind\,Ab is the nearest cold-Jupiter system to Earth. Asteroseismology of $\epsilon$\,Indi\,A has revealed a frequency of maximum power, $\nu_{\rm max}$, of 5265\,$\upmu$Hz -- the highest so far of any solar-like oscillator -- which in turn allows its mass to be inferred at $M=0.782$\,M$_{\odot}$ \citep{lundkvistetal2024}. It also has an asteroseismic large spacing, $\Delta\nu$, of 201.25\,$\upmu$Hz, which corresponds to an age younger than 4\,Gyr \citep{campanteetal2024}. These asteroseismic measurements not only verify the distance to the system independent of parallax, but also provide important constraints on the brown dwarf properties, which are the only examples well constrained and nearby enough to have allowed a test of substellar cooling with time for a T~dwarf. Thus the $\epsilon$\,Indi system is a textbook example of the powerful synergy of asteroseismology and binary star analyses.

\section{Solar-like oscillators beyond the main sequence}
\label{sec:rgs}

We have focussed so far on systems currently having companions, but as stars evolve up the red giant branch (RGB) and their radii expand, they can become common envelope binaries and eventually mergers \citep{paczynski1976}. From the outside, it is impossible to distinguish a single 2.5-M$_{\odot}$ star that has evolved to become a red giant, from a 1.5 + 1.0-M$_{\odot}$ pair that merged on the RGB. However, their oscillation spectra can tell them apart. The Brunt-V\"ais\"al\"a frequency of the merger product looks similar to the original 1.5-M$_{\odot}$ primary, and significantly different from a 2.5-M$_{\odot}$ single star of the same luminosity \citep{rui&fuller2021}.

In the age of space photometry, red giants are often characterised by their period spacings, $\Delta\Pi$, and their frequency spacings, $\Delta\nu$. Specifically, it is common to plot many stars in the $\Delta\Pi$--$\Delta\nu$ plane, where RGB stars fall on a well-defined sequence. \citet{deheuvelsetal2022} showed that merger products can be identified in this diagram. Ordinarily, intermediate-mass red giants, which reach the core-helium burning phase with non-degenerate cores, lie above the RGB sequence in this diagram. But if those stars attained their mass by merger, especially after one of those stars developed a degenerate core, they instead lie under the RGB sequence. Hence, it is only the relatively uncommon case where two stars merge on the main sequence that leaves no asteroseismic fingerprints.

Binaries can also be detected indirectly before they merge, via evidence of substantial mass loss by Roche lobe overflow (RLOF) onto a companion. Ordinarily, stars might lose up to 0.2\,M$_{\odot}$ through wind-driven mass loss as they ascend the red giant branch. Then, when they begin core helium burning, they settle onto a line called the zero-age helium burning (ZAHeB) line that is sharp in $\nu_{\rm max}$ both observationally and in models. (This is entirely analogous to the ZAMS forming a well-defined lower edge to the main sequence in luminosity space.) \citet{ylietal2022} found a handful of stars that fell to the wrong side of the ZAHeB line. They showed that these stars must have had non-degenerate cores (i.e. high main-sequence masses $\gtrsim2.2$\,M$_{\odot}$) when they ascended the RGB, but their observed asteroseismic quantities ($\Delta\nu$ and $\nu_{\rm max}$) indicate their masses are now much lower ($\lesssim1.6$\,M$_{\odot}$). These stars therefore lost substantial mass -- much more than is explained via wind-driven mass loss -- suggesting they donated it to an otherwise undetected companion.

Other tell-tale signs of significant mass loss are conspicuously low masses on the RGB. The 13.8-Gyr age of the Universe puts a lower limit on the mass of an RGB star of approximately 0.8--1.0\,M$_{\odot}$, yet \citet{ylietal2022} discovered dozens of red giants with masses significantly lower than this limit. Since ``stellar winds driven by radiation and pulsation can only remove up to 0.2\,M$_{\odot}$,'' they argue, those stars must have undergone more extreme mass loss, with RLOF onto a companion being the prime culprit. A specific example concerns the low-mass, core helium burning star KIC\,4937011, which is a member of the cluster NGC\,6819. This $\sim$2.4-Gyr-old cluster has a main-sequence turn-off mass of $\sim$1.6\,M$_{\odot}$, yet the asteroseismic mass of KIC\,4937011 is 0.74\,M$_{\odot}$ \citep{matteuzzietal2024}. Not only is this star too low mass to be a red giant given the age of the cluster, the same is true given the age of the universe. \citet{matteuzzietal2024} hypothesise that KIC\,4937011 has lost around 1.0\,M$_{\odot}$ of material through common envelope ejection in its recent history.

Of course, not all systems destined to experience common envelopes or RLOF have yet begun to lose mass. If a star does not overflow its Roche lobe at the tip of the RGB, it will have another chance as it ascends the asymptotic giant branch (AGB). This is the fate of any core-helium-burning red giants with companions in orbits of $200 < P_{\rm orb} {\rm (d)} \lesssim 3000$ \citep{oomenetal2018}, which includes several hundred known binaries with oscillating RG components \citep{themessletal2018,becketal2024}. In some cases, the pulsations of the companion to the red giant are also detectable, and in some of {\it those} cases, those pulsations can be used to derive an orbital solution. That is the subject of the next section.

\section{Pulsation timing orbits}
\label{sec:TDs}

The orbits of binary systems can be obtained by timing the pulsations of one (or more) components in the system. In its simplest form, the method is the classical O-C method \citep{barnes&moffett1975,sterken2005c}, which has been adapted to readily combine signals from many different pulsation modes \citep{murphyetal2014}. There are also implementations analogous to frequency-modulated radios, where the stellar pulsations are the carrier frequencies modulated by the binary motion \citep{shibahashi&kurtz2012,shibahashietal2015}. These variations on a methodological theme have been confirmed with other techniques, such as RVs and eclipses \citep{barlowetal2011a,barlowetal2011b,kurtzetal2015a}. The 2018 review \citep{murphy2018} covered pulsation timing in detail, most of which will not be repeated here. Instead, the following paragraphs will cover the advancements from 2018, what can be expected in the near future, and will conclude with another demonstration of the synergy of pulsations and binary analyses.

As of 2018, a few hundred pulsation timing binaries were known, most of which contain one or more $\delta$\,Sct pulsators \citep{murphyetal2018}\footnote{Some $21\pm6$\% of these $\delta$\,Sct stars are not the original primaries of their systems, and have {\it received} mass through RLOF in their binary evolution}. Other classes of variables are not expected to yield many pulsation-timing binaries \citep{comptonetal2016,hermes2018}, yet that number is not zero. For instance, \citet{heyetal2019} applied the technique to six rapidly oscillating Ap stars, yielding two candidate (but unconfirmed) binaries. \citet{otanietal2022} derived a convincing pulsation timing orbit from the subdwarf B star in AQ Col, but caution should generally be applied when dealing with subdwarfs because their modes undergo sporadic period changes \citep{mackebrandtetal2020}. It is occasionally useful, of course, to be able to use pulsation timing to know when a star is {\it not} in a binary, too \citep{szewczuk&dd2018,murphyetal2020d}. This has been applied to rule out stellar companions as false positives in transiting exoplanet searches \citep{heyetal2021}.

The precision of orbital solutions can be greatly enhanced by combining pulsation timing with other methods. For instance, radial velocity is the time-derivative of the light arrival-time delay probed by pulsation timing, and the complementarity of the mathematical functions yields solutions more precise than doubling the precision of either method alone (e.g. \citealt{murphyetal2016b,derekasetal2019,lampensetal2021}). These two methods both operate along the radial projection of the orbit, hence ultimately confer the same information. The next frontier was to incorporate astrometry data into the analysis, as the astrometric plane is perpendicular to the radial vector.

The only analysis to have combined pulsation timing and astrometry is that of the $\alpha$\,Pic system by \citet{dholakiaetal2024}. It had long been suspected that $\alpha$\,Pic was binary, but neither RVs nor astrometry could adequately sample its 1316-d orbit. Using three pulsation modes of $\alpha$\,Pic\,A, \citet{dholakiaetal2024} used TESS data to measure variations in light arrival times, constraining the orbital period, and jointly fitted astrometric data from Hipparcos. The addition of astrometry allowed an orbital inclination to be inferred in addition to the usual orbital parameters from radial analyses, and further parameters such as a mass ratio then follow from comparison of the photocentre motion to the amplitude of the light arrival-time delays. Many more joint astrometric solutions can be expected following Gaia DR4, when intermediate astrometric data are released.

Most pulsation timing studies conducted thus far have been based on \textit{Kepler} data, with a few TESS systems having been investigated ad-hoc (e.g. \citealt{yangetal2021b,vaulatoetal2022}). The use of multi-mission photometry is also on the rise \citep{xin&ming2023,ziebaetal2024}. The first systematic pulsation timing study on TESS photometry has now been conducted \citep{dholakiaetal2024}, which used data from the first four years, corresponding to TESS Sectors 1--55. A disadvantage of TESS photometry compared to \textit{Kepler} is the gaps in coverage due to TESS's step-and-stare observing strategy \citep{rickeretal2015}. \citet{dholakiaetal2024} therefore also evaluated the minimum number of Sectors of observation required to reliably infer the orbital parameters. As TESS continues observing, the number of systems discovered and the quality of individual orbits are both expected to increase. In 2025, the launch of PLATO will provide yet another source of high duty cycle light curves for such analyses.

Improvements to pulsation timing are not limited to the inclusion of more data. In fact, some developments have focussed on better use of the data at hand. A drawback of the original method was the need to subdivide the time-series to look for pulsational phase shifts in each subdivision. This can be obviated by forward-modelling the light arrival-time delays onto the light curve itself. \citet{heyetal2020a,heyetal2020b} built and provided the tools to apply this, which could be extended to include other parametric phenomena such as eclipse models in future \citep{prsaetal2022}. Dispensing with the need to subdivide the light curve is particularly advantageous when probing small orbits, which naturally includes most of the EB population. \citet{murphyetal2016b} had to introduce corrections for undersampling of the orbits when subdividing, but there is no such need when forward modelling.

Yet more discoveries abound in the judicious application of pulsation timing to other $\textit{Kepler}$ data sets. Simply by extending the search space beyond the $T_{\rm eff}$ range of the \citet{murphyetal2018} catalogue, new systems have been found in \textit{Kepler} data \citep{manzoori2020,yang2022,murphyetal2020a}. And use of the collateral `smear' data on the \textit{Kepler} CCD to produce light curves for otherwise saturated targets has yielded even more \citep{popeetal2019}. A systematic application of pulsation timing to eclipsing systems is yet to be undertaken and is an obvious future direction.

One specific system exemplifies the combined power of asteroseismology and binary studies, and also happens to have been too cool (KIC $T_{\rm eff}$ = 5406\,K) to have been captured in the primary pulsation-timing survey. KIC\,9773821 is a RG--$\delta$\,Sct binary in which both components have oscillations detected with \textit{Kepler}, i.e. it is a PB2. The $\delta$\,Sct pulsations confer a pulsation timing orbit, but the stochastic nature of the RG oscillations do not. The sharp spectral lines of the RG permit an RV orbit, while the broad $\delta$\,Sct lines do not. Studied together, these allow a joint orbital model including a mass ratio \citep{murphyetal2021b}. Asteroseismology of the RG gives its mass, which then also gives the $\delta$\,Sct mass via the orbital mass ratio. However, that RG mass is 2.1\,M$_{\odot}$, so the RG core was never degenerate. This made it difficult to tell if the RG was core He burning, or on the RGB instead, hence its age was uncertain. This dilemma was solved using the $\delta$\,Sct star -- the RGB scenario would have required the $\delta$\,Sct star to be somewhat younger and hotter; so hot, in fact, that it would lie outside the $\delta$\,Sct instability strip. Although this is not unheard of (\citealt{bowman&kurtz2018,murphyetal2019,sharmaetal2022,hey&aerts2024,mombargetal2024a}, cf. \citealt{kahramanalicavusetal2024}),\footnote{\citealt{sharmaetal2022}, in particular, cited the many investigations into this.} this argument was one of many indicating that the primary had to be core helium burning, and illustrates the benefit of combining asteroseismic and binary analyses.

\section{Tidally trapped pulsations}
\label{sec:tidal}

Some orbits are too small for pulsation timing to be effective, even with the latest methods. This includes an entirely new class of pulsating binaries -- tidally trapped pulsators. These are a subset of the broader class of tidally perturbed pulsators, where the two components of the binary are sufficiently close that tidal forces affect the pulsations. First discovered in \textit{Kepler} data \citep{hambletonetal2013,balona2018b}, these systems sprang to the fore with TESS, starting with U\,Gru \citep{bowmanetal2019}, and several studies followed shortly thereafter \citep{lee2021,southworth2021,steindletal2021a,vanreethetal2022}. The first systematic search for such systems produced five new discoveries in a well-referenced study by \citet{vanreethetal2023}.

The hallmark of tidally trapped pulsations are pulsation amplitudes that vary strongly with orbital phase and in addition to the effects of eclipses masking the pulsator. The amplitude variation results from tidally-induced asphericity, and \citet{fulleretal2020} described the three resulting phenomena. The first is tidal alignment. Ordinarily the stellar rotation axis is the axis of symmetry for pulsations. Not so in close binaries, where tidal braking of the stellar rotation \citep{tassoul&tassoul1992a,tassoul&tassoul1992b} and tidal distortion lead to the tidal axis being the axis of alignment instead. Hence, the pulsation pole comes into and out of view twice each orbital cycle. The second phenomenon is tidal trapping, wherein the modes themselves are spatially confined to either the tidal poles or the tidal equator, further affecting their visibility over an orbital cycle. The third phenomenon is tidal amplification. This effect occurs when modes are able to propagate closer to the stellar surface. A pulsation wave's outer turning point is defined by the acoustic cutoff frequency, which depends on the surface gravity and is ordinarily latitude independent. But in a close binary, the surface gravity is markedly lower at the L1 Lagrangian point, allowing the modes to propagate closer to the surface and become `amplified'.

More work on tidally trapped pulsators can be expected in the near future as different geometries are observed for the first time (e.g. systems viewed from the tidal poles), and as modellers keep pace with the rate of observations to include the effects of tides on the pulsation models.

\section{Pulsators in eclipsing binaries}
\label{sec:EBs}

The proliferation of space-based light curves, particularly resulting from TESS's all-sky coverage, has dramatically increased the number of known eclipsing binaries \citep{prsaetal2022}. This is particularly the case for B stars, which were not observed in large numbers by \textit{Kepler} due the avoidance of bright stars and due to the \textit{Kepler} field lying out of the galactic plane \citep{kochetal2010}. TESS's smaller size lends itself to observing brighter stars than \textit{Kepler} and has indirectly resulted in many of the newly-discovered EBs being A stars, which strike a balance between being bright and numerous (\citealt{ijspeertetal2021}; Southworth, these proceedings). It is no surprise, then, that the number of studies on pulsating stars in EBs suddenly shot up in the TESS era \citep{southworthetal2020,buddingetal2021,lee&hong2021,southworthetal2021,erdemetal2022}. While most of these `early' studies concerned single systems, it was not long before ensembles were also collated \citep{southworth&bowman2022,kahramanalicavusetal2022,kahramanalicavusetal2023,ulacs&ulusoy2023,catanzaroetal2024,eze&handler2024}.

The talk upon which this review is based had concluded with a couple of lessons learned by the author from recent studies of pulsators in EBs, and so this review concludes the same way. The first of those lessons was that spin-orbit alignment of binary systems cannot always be assumed. This lesson came from the VV\,Ori system, consisting of a $\beta$\,Cep pulsator and an SPB star in a 1.49-d orbit, recently reported to show changes in the primary's inclination and changes in the totality of its eclipses \citep{southworthetal2021,buddingetal2024}. Earlier analyses of many other systems as part of the BANANA project (`Binaries Are Not Always Neatly Aligned') held the same clues, wherein systems with good alignment and others with misalignment were found \citep{albrechtetal2009,albrechtetal2011,albrechtetal2012,albrechtetal2014}.

The second lesson came from the triple system V1031\,Ori, which has a $\delta$\,Sct component. That lesson was that Gaia distances for binaries and multiples are not yet trustworthy (as of Gaia DR3). The distance to this system from the EB analysis has yielded $210\pm26$\,pc \citep{lee2021}, which compares favourably to earlier comprehensive studies ($215\pm25$\,pc, \citealt{andersenetal1990}) and Hipparcos ($205\pm30$\,pc, \citealt{vanleeuwen2007}). However, it is somewhat discrepant with the distance from Gaia EDR3 [$588^{+389}_{-198}$\,pc (geometric) and $734^{+290}_{-258}$\,pc (photogeometric); see \citealt{lee2021} for details]. Even in the full DR3 release, the {\tt gspphot} distance is inaccurate at $150.8^{+3.3}_{-3.6}$\,pc, though it is now underestimated rather than overestimated. The problem is not universal -- HD\,23642 is another EB with a $\delta$\,Sct component and has good agreement between the distance derived from eclipses ($134.7\pm2$\,pc, \citealt{southworthetal2023}) and from Gaia ({\tt gspphot} $=137.6\pm1$\,pc in DR3). Pending updated astrometry in Gaia DR4, V1031\,Ori serves as a caution.

\acknowledgements
SJM was supported by the Australian Research Council (ARC) through Future Fellowship FT210100485.

\bibliography{sjm_bibliography}

\end{document}